\documentclass[12pt]{article}
\usepackage{amsfonts}

\usepackage{amsmath}
\usepackage{graphicx}

\input{tcilatex}

\begin{document}

\title{Anomalies of the tides}
\author{Thomas H. Otway \\
\textit{Department of Mathematics,} \textit{Yeshiva University,}\\
\textit{500 W 185th Street, New York, NY \ 10033}}
\date{}
\maketitle

\begin{abstract}
\textit{\ }Unusual features of water waves that Galileo described in a
letter to Cardinal Orsini in 1616 are revisited from the perspectives of
singular optics and geometric analysis.
\end{abstract}

{\large Contents}

\bigskip

1. Introduction: Extreme variants of tides, waves, and currents

2. Optical analogies

\qquad2.1. Caustics

\qquad2.2. Wave dislocations

3. Tidal bores

\qquad3.1. The wave prior to the formation of the caustic

\qquad\qquad3.1.1. The applicability of geometrical optics to the tides

\qquad3.2. The wave after formation of the caustic

\qquad\qquad3.2.1. The steady flow of shallow water

4. Waves in a neighborhood of a caustic

\qquad4.1. Lagrangian manifolds

\qquad4.2. The caustic boundary layer

\qquad\qquad4.2.1. Singularities which are artifacts of the linearization
method

\qquad4.3. Rogue waves

5. Amphidromic points

\qquad5.1. Wavefronts

Notes

References

\section{Introduction: Extreme variants of tides, waves, and currents}

\textit{``These conflicting motions, depending on the different positions
and lengths of the interconnected seas and on their different depths, give
rise sometimes to those irregular disturbances of the water whose causes
have worried and continue to worry sailors, who experience them without
seeing any winds or other atmospheric disturbances that might produce them.''%
}

Galileo Galilei, to Alessandro Cardinal Orsini$^{1}$

\bigskip

The primary purpose of Galileo's theory of the tides had nothing to do with
the sea. He considered it his most ``stringent'' proof of the motion of the
Earth about the Sun. And yet, his letter to Cardinal Orsini of 1616
contained a host of observational details about variations and anomalies of
tides and currents, and it is easy to believe that the tides interested
Galileo for their own sake as well.

In this review we consider issues raised by Galileo concerning variations in
tides over the known world of his time, and investigate the current status
of similar issues.

In one sense, all such matters were resolved by the end of the $17^{th}$
century with the advent of Newtonian dynamics. The gravitational force
exerted by a celestial body on the waters of the Earth is proportional to
the mass of the body divided by the square of its distance.$^{2}$ It is easy
to compute that the mean gravitational force exerted by the Sun on a
particle of seawater is much greater than that exerted by the Moon. But
tides are not due to the mean force exerted by a heavenly body, but to the
variation of that force as the Earth rotates on its axis. Because of the
small distance between the Earth and the Moon relative to the Earth's
average radius, the deviation of the Moon's mean gravitational attraction
over a 24-hour period greatly exceeds that of the Sun, and one expects that
the tides will vary most noticeably with the position of the Moon relative
to a fixed point on the Earth. In particular, one is led to expect a high
tide when the Moon is overhead and another high tide when the Moon is on the
other side of the Earth. Taking viscous effects involving friction with the
ocean floor into account, this is essentially what is observed. The effect
of the Sun's gravitational field can be measured in \textit{spring} tides,
when the Sun is aligned with the Earth and the Moon, and \textit{neap}
tides, when the three bodies form a right angle. The elliptical orbit of the
Moon produces further variations in tidal amplitude, increasing by about
20\% the height of the tide at lunar \textit{perigee} (nearest approach). In
this way, Newton's \textit{equilibrium theory} of the tides explains the
periodic vertical displacements of the oceans.$^{3}$ Horizontal
displacements, or \textit{tidal currents,} were derived by Laplace almost a
hundred years later, under strong simplifying assumptions, in his \textit{%
dynamical theory} of the tides.

But Galileo's letter includes a long passage on \textit{``the properties of
tides observed in experience.''}$^{4}$ These properties are not easily
derived from first principles. Galileo notes the absence of observable tides
in certain bodies of water, including some seas, as well as geographical
variations in tidal frequencies and the apparent existence of tide-induced
currents. He also reports geographic variations in tidal amplitudes,
including geographic variation within a single body of water and among
certain straits. It is the variation in tidal amplitudes that we consider
here.

The equations governing the propagation of water waves are highly nonlinear,
and solutions can be expected to be sensitive to even small variations in
boundary conditions. For this reason most variations in tidal amplitude are
ascribed to local boundary geometry, such as the shape of a continental
shelf or ocean basin, or to the influence of local currents. But others are
of more subtle origin. We consider two kinds of \textit{anomalies,} extreme
variants, which seem to arise from singularities of the tidal potential, and
which thus reflect topological rather than geometric conditions. One class
of anomalies consists of \textit{tidal bores,} which are studied in Sec. 3.
These take the form of great solitary channel waves, traveling at constant
speed and constant elevation. The other class consists of \textit{%
amphidromic points,} places in the ocean where the tides disappear entirely;
these are studied in Sec. 5.

Despite references to \textit{``immense currents''} in narrow channels and
to seas \textit{``in a state of exhaustion,''} neither tidal bores nor
amphidromic points are specifically mentioned in Galileo's letter. A
bore-like solitary wave created by a horse-drawn barge was witnessed by J.
Scott Russell in 1834 and described by Russell in 1844; amphidromic points
of the North Sea were described by William Whewell in 1833 and 1836. These
appear to have been the first scientific descriptions of the two classes of
anomalies.$^{5}$

Galileo was led by his erroneous theory of the tides to consider
hydrodynamic events which have no direct relation to the gravitational
attraction of the Moon, whose \textit{``dominion over the waters''} he in
any case scornfully rejected. (Galileo considered the notion of lunar
influence on the tides to be not only wrong but puerile.$^{6}$) Rather,
Galileo believed that the tides represented the sloshing of water due to
combined effects of the axial rotation and orbital revolution of the Earth.
He correctly suggested that many tidal anomalies arise from geographical
effects involving the width, depth, and geometry of the particular bodies of
water and their relation to contiguous bodies of water. Moreover, apparent
Coriolis acceleration, due to the axial rotation of the Earth, does modify
certain tidal currents, which is somewhat in the spirit of Galileo's
argument. (See also endnote 50.) That the orbital revolution of the Earth
cannot have any effect on the tides was shown by Daniel Bernoulli in 1752.
The main thrust of Galileo's theory is thus completely wrong.$^{7}$ The
examples cited by Galileo in defense of his theory lead us in Sec. 4 to
consider the general topic of water-wave singularities, which are manifested
as tidal anomalies, rogue waves, and other remarkable events.

\section{Optical analogies}

Anomalies of the tides and other ocean waves can be viewed as distant
relatives of singularities that occur in both geometrical and physical
optics. Although the research field called \textit{singular optics} is
rather new,$^{8}$ the study of singularities in families of light rays goes
back certainly to Galileo's century, if not to his own lifetime.
Singularities in the ray theory, known as \textit{caustics}, enter the
theory of water waves in the form of tidal bores and, more recently, models
of rogue waves. Phase singularities, known as \textit{dislocations}, enter
the theory of water waves in the form of amphidromic points. The study of
phase singularities goes back, in a sense, to the work of Grimaldi and
Newton in the $17^{th}$ and early $18^{th}$ centuries.$^{9}$

\subsection{Caustics}

\textit{Optical caustics} are points, curves, or surfaces of anomalously
bright light, often observed as a double arc of reflected sunlight on the
surface of a cup of coffee. Caustics may also appear as a web of wavy lines,
shimmering on the floor of a swimming pool on sunny days or reflected on the
hull of a boat. Caustics can be created by reflection, as in the coffee-cup
caustic, or by refraction, as in swimming-pool caustics. Caustics created by
reflection are called \textit{catacaustics}, and those created by
refraction, \textit{diacaustics}.

Caustics have a geometric representation as the \textit{envelope} of a
family of light rays, that is, as curves which share a common tangent with
every member of the family of rays. (In higher dimensions these are called 
\textit{developable surfaces}.) In order for the geometric representation to
acquire physical meaning, it is necessary to imagine that light propagates
in discrete quanta which travel along straight paths. In fact, the
propagation of light in most circumstances is well-approximated by the
motion of a wave. Wave motion implies diffractive effects. The human eye has
evolved in such a way that the zero-wavelength approximation, in which light
can indeed be treated as a particle traveling along a straight path, is
adequate for interpreting most optical phenomena of everyday life. This
approximation is known as \textit{geometrical optics}.

Caustics were studied intensively in the $17^{th}$ century and great
discoveries were made about them. Descartes correctly interpreted the ray
structure of the rainbow 1637. His argument by ray tracing was sufficiently
accurate to reproduce the angle at which this caustic is visible, but could
not account for dark curves called \textit{superluminaries}; these were
explained by Airy in the $19^{th}$ century, applying a model that took into
account diffractive effects. By 1678 Huygens had solved the catacaustic
problem for a concave circular reflector and a \textit{pencil} of rays
emanating from a single light source at infinity. He obtained the nephroid
as the catacaustic curve, which explains the coffee-cup caustic. Huygens'
result seems to have been independently discovered by Tschirnhaus, a
mathematician who achieved wider fame as a maker of porcelain.$^{10}$ The
cycloidal caustic arising from a light source on a circle was discovered by
Jacob and Johann Bernoulli in 1692. Although both Airy and Cayley studied
caustics in the mid-$19^{th}$ century, interest in them generally lapsed
between the $17^{th}$ century and the latter half of the $20^{th},$ when
they were rediscovered as beautiful examples of the singularities that arise
in differentiable mappings.$^{11}$

\subsection{Wave dislocations}

In the geometrical optics approximation, the caustic is a true singularity
in the sense that energy at a caustic tends to infinity. In fact diffractive
effects, which are ignored in geometrical optics, keep the actual energy
finite at a caustic. The diffraction of light was also discovered in the $%
17^{th}$ century, in the experiments of Grimaldi and in the optical wave
theory of Huygens.$^{12}$ It is reasonable to ask whether caustics, so
strongly identified with ray optics, have an analogy in the study of wave
optics.

Of course singularities need not involve infinite quantities. They may, for
example, involve indeterminate quantities. The propagation of light can
often be represented by a complex wave having the form 
\begin{equation*}
\psi\left( \mathbf{r},t\right) =\func{Re}\left\{ \alpha\left( \mathbf{r}%
,t\right) \exp\left[ i\chi\left( \mathbf{r},t\right) \right] \right\} , 
\end{equation*}
where the amplitude $\alpha$ and phase $\chi$ are real scalar functions, $t $
is time, $\mathbf{r}$ is the radial vector, and $\func{Re}$ denotes the real
part of a complex function; here and throughout, $i^{2}=-1.$ In this case
the quantity $\chi\left( \mathbf{r},t\right) $ is indeterminate at the zeros
of $\alpha\left( \mathbf{r},t\right) .$ The indeterminacy has a physical
manifestation as a singularity of the wavefront. Such \textit{dislocations}
of light waves are sometime called \textit{optical vortices}, for reasons
that will become apparent in Sec. 5.1, where we discuss dislocations of
water waves.

If optical caustics are observed as especially bright regions, optical wave
dislocations will be observed as especially dark regions. Berry has called
wave dislocations ``singularities of faint light,'' and has observed$^{13}$
that conditions under which caustics are easily seen are those under which
dislocations, which are sub-wavelength fine structure, tend to be invisible,
whereas high-magnification conditions under which dislocations are clearly
visible are those under which diffractive effects tend to smear out the
caustics.

\section{Tidal bores}

On the largest scale, the theory of the tides is a linear theory. Under
appropriate simplifying assumptions, both the horizontal current and the
vertical surface elevation satisfy the two-dimensional wave equation with
propagation speed $\sqrt{gh},$ where $g$ is the acceleration due to gravity
and $h$ is the depth of the ocean floor. These simplifying assumptions are,
essentially, small wave amplitude and shallow depth.$^{14}$ Due to their
enormous wavelength $-$ reaching, in the absence of disturbing land masses,
half the circumference of the Earth $-$ the ocean tides satisfy both
assumptions and can be treated as an approximately linear system.

But the linear wave approximation is too rough to account for the amplitude
singularities which concern us; the singularities are thrown out with the
neglected nonlinear terms. To better account for local deviations we retain
the shallow-water hypothesis only. We recover singular solutions, but at the
expense of also recovering some of the nonlinearity.

It was noticed as early as the $19^{th}$ century that water entering a
narrow channel is sometimes funneled by the geometry of the entrance in such
a way that the incoming current creates a pressure wave. This model yields a
qualitative explanation of, for instance, the well known bore observed on
the Tsien-Tang River in northern China, a steep and turbulent solitary wave
caused by the rising tide pushing water into a narrow estuary.

Alternatively, we can imagine that a narrowing at the mouth of a channel 
\textit{focuses} tidal waters by a inducing a change in their velocity as
the tide enters the still water of the channel. A similar effect occurs as a
wave progresses up a sloping beach. In each case the change of velocity
corresponds to the focusing effects in optics which arise from changes in
the refractive index of the medium. This natural focusing of the incoming
tide is analogous to the production of a caustic at a ray singularity when
the Sun's rays are focused through the lens of a magnifying glass.

\subsection{The wave prior to the formation of the caustic}

We consider a two-dimensional, inviscid, incompressible, irrotational flow.
The first assumption means that we take a vertical cross section of the
ocean and consider the boundary curve representing the vertical displacement
of the water. (A different interpretation of two-dimensional flow will be
given in Sec. 3.2.1.) The flow of water is well-approximated as the movement
of \textit{lamina}, or arbitrarily thin horizontal layers. In ignoring
viscosity, we ignore \textit{shear}, or tangential forces between the
layers, the fluid equivalent of friction. By \textit{incompressibility}, we
mean that the density of the medium is taken to be constant, and by \textit{%
irrotationality,} that a massless horizontal paddle wheel placed in the flow
may move, but will not rotate. Mathematically, this last assumption is
equivalent to a vanishing curl of the velocity vector.

In writing the equations of motion we take the $x$-axis to represent the
undisturbed surface of the water. The undisturbed water depth is represented
by the curve $y=-h(x),$ where $h$ is a function having nonnegative range.
Forces acting on the water cause the surface to deviate from the undisturbed
condition represented by the line $y=0.$ These time-dependent deformations
of the water surface are given by the family of curves $y=\eta\left(
x,t\right) .$ We say that the surface is \textit{free,} which is to say that
the function $\eta\left( x,t\right) $ is unknown and must be determined as
part of the solution of the equations.

Denote by $u(x,y,t)$ the component of the velocity in the $x$-direction and
by $v(x,y,t)$ the component of velocity in the $y$-direction. Writing the
velocity components in terms of the velocity potential $\varphi\left(
x,y,t\right) ,$ we have the local relations 
\begin{equation*}
u=\frac{\partial\varphi}{\partial x},\;v=\frac{\partial\varphi}{\partial y} 
\end{equation*}
for irrotational flow, as well as the \textit{continuity equation} 
\begin{equation*}
\frac{\partial^{2}\varphi}{\partial x^{2}}+\frac{\partial^{2}\varphi}{%
\partial y^{2}}=0 
\end{equation*}
for incompressible flow, where $x\in\mathbb{R}$ and $y\in\left[
-h(x),\eta\left( x,t\right) \right] .$ The continuity equation must be given
boundary conditions in order to have a unique solution. In particular, it is
natural to assume that the derivative of $\varphi$ in the normal direction
vanishes at the depth $y=-h(x).$

Other constraints are imposed by the physics. These are mathematically
necessary in that $\eta$ as well as $\varphi$ must be determined uniquely.
The pressure $p$ associated to the flow is governed by \textit{Bernoulli's
formula} 
\begin{equation*}
\frac{p}{\rho}+\frac{Q}{2}+gy+\frac{\partial\varphi}{\partial t}=C, 
\end{equation*}
where $C$ may depend on $t$ but is independent of $x$ and $y,$ $\rho$ is
mass density, and 
\begin{equation*}
Q\equiv u^{2}+v^{2}. 
\end{equation*}
(Note that the constant $C$ can be taken to be any convenient number by
applying a\textit{\ gauge transformation} of the form 
\begin{equation*}
\widetilde{\varphi}=\varphi-\int^{t}C\left( \tau\right) d\tau+kt 
\end{equation*}
for an appropriately chosen constant $k.$ However, in Sec. 3.2.1 we will
apply Bernoulli's equation to the case of steady flow, in which $\partial
\varphi/\partial t=0.$ That assumption fixes the value of $C.$) In addition,
we have a \textit{dynamic boundary condition} on the free surface pressure, 
\begin{equation*}
p_{|y=\eta}=0, 
\end{equation*}
and a \textit{kinematic boundary condition,} 
\begin{equation*}
v=\frac{\partial\eta}{\partial t}+u\frac{\partial\eta}{\partial x}, 
\end{equation*}
governing velocity components at the free surface. The kinematic condition
restricts the motion of surface particles to tangential directions, so that
surface particles remain on the surface.

The exact nonlinear theory just presented is too hard to solve analytically.$%
^{15}$ A useful approximation results from adding a single physical
assumption, that the pressure is well-approximated by the \textit{%
hydrostatic law} 
\begin{equation*}
p=g\rho\left( \eta-y\right) . 
\end{equation*}
Physically, this amounts to the hypothesis that the $y$-component of
acceleration of the water particles has a negligible effect on pressure, a
hypothesis which would tend to be satisfied at shallow depths.

For example, in the simple case in which the ocean floor has constant slope $%
m,$ it can be shown$^{16}$ that the equations of motion imply the system 
\begin{equation}
\left\{ \frac{\partial}{\partial t}+\left( u+c\right) \frac{\partial }{%
\partial x}\right\} \left( u+2c-mt\right) =0,
\end{equation}
\begin{equation}
\left\{ \frac{\partial}{\partial t}+\left( u-c\right) \frac{\partial }{%
\partial x}\right\} \left( u-2c-mt\right) =0,
\end{equation}
where $m=dh/dx$ and 
\begin{equation*}
c\left( x,t\right) =\sqrt{g\left( \eta+h\right) } 
\end{equation*}
is the local speed at which small wave-like disturbances advance relative to
the water. In physical terms this system asserts that the function $u\pm2c-mt
$ is constant for any point that moves through the water at speed 
\begin{equation*}
\frac{dx}{dt}=u\pm c. 
\end{equation*}
This gives two families of so-called \textit{characteristic curves} for eqs.
(1) and (2).

The characteristic curves simplify if the channel floor is assumed
horizontal $(h=constant)$ and if at $t=0,$ $\eta=0$ and we have $%
u=u_{0}=constant$ and $c=c_{0}=\sqrt{gh}.$ We assume that a disturbance is
initiated at $x=0.$ In this case one of the two families of characteristic
curves will reduce to a family of straight lines, along which both $u$ and $%
c $ will have constant value.

Letting the curve $x=x(t)$ in the $xt$-plane denote the displacement of
water particles initially lying in a vertical plane at $(x,t)=(0,0),$ one
finds that the slope of this curve is given by the differential equation 
\begin{equation*}
\frac{dx}{dt}=\frac{3}{2}u_{A}+c_{0}, 
\end{equation*}
where $u_{A}$ is the velocity of the incoming tide. In cases for which $%
u_{A} $ increases with $t,$ the resulting characteristics will intersect at
some value of $t$ and will, in general, possess an envelope. In the language
of Sec. 2, the envelope is a caustic of the family of rays formed by the
straight-line characteristics of eqs. (1), (2).

In the limiting case in which the curve $x(t)$ representing the action of
the lens is a straight line issuing from the origin, discontinuity solutions
can be found analytically.$^{17}$ Moreover, the shallow-water equations can
be linearized by a \textit{hodograph transformation,} in which the
independent and dependent variables of the equations are switched.
Precisely, define new variables 
\begin{equation*}
\xi=u-2c-mt, 
\end{equation*}
\begin{equation*}
\sigma=u+2c-mt 
\end{equation*}
and let $\left( x,t\right) \rightarrow\left( \xi,\sigma\right) .$ This
hodograph transformation, followed by an additional differentiation,
converts equations of the form (1), (2) into a cylindrical wave equation.$%
^{18}$

\subsubsection{Technical note: The applicability of geometrical optics to
the tides}

One might object to the use of caustics in analyzing singularities of the
tides. Caustics are objects that arise in geometrical optics. Thus they are
high-frequency phenomena, whereas the tides are probably the
lowest-frequency waves encountered in everyday experience.

This objection can be met on both mathematical and physical grounds. The
caustics which appear in the preceding argument are envelopes of
characteristic lines associated to a partial differential equation.
Characteristic curves can be associated to any differential equation of
hyperbolic type, regardless of physical context. From a physical point of
view, the relevant dimension for applying the geometrical optics
approximation is the ratio of the width of the channel entrance to the
horizontal scale of the incoming tide. In cases for which this ratio is
small, ray effects will dominate over diffractive effects at the mouth of
the channel.

\subsection{The wave after formation of the caustic}

We mentioned that tidal bores can be produced in a channel by refraction of
the advancing tide due to the narrowing of the width of the channel. In this
case it may be possible to assume, as we have in part of Sec. 3.1, that the
channel depth is constant. Refraction can also occur in a channel of
approximately constant width due to the effect of a sloping channel floor.
The mechanism of bore formation in this case is analogous to the formation
of a breaking wave on a sloping beach. In this section we assume that the
channel depth decreases by a vertical distance $\delta$ over a small spatial
interval. Just as a sudden narrowing of the channel appears to be the cause
of the bore on the Chinese Tsien-Tang River, a sudden decrease in channel
depth appears to explain the formation of tidal bores on the English rivers
Severn and Trent.$^{19}$

The development of a tidal bore cannot be modeled by steady flow, which is
by definition time-independent. But it is not unreasonable to suggest that
the velocity of a propagating bore at any fixed point of the channel is
constant over the time interval of interest. We will assume this, and we
will also assume that both $\delta$ and the height $\varepsilon$ of the
advancing bore above the height $H$ of the undisturbed water are
sufficiently small. Under these assumptions one can derive a remarkably
simple model for the evolution of a tidal bore once it has formed.$^{20}$

Consider a steady current flowing in the positive-$x$ direction. Suppose
that an incline of magnitude $\delta$ occurring between the points $x_{0}$
and $x_{1}$ of an otherwise horizontal channel floor produces through
refraction an elevation of height $\varepsilon$ in the surface of the water
at $x=x_{1}.$ Suppose that $x_{0}<x_{1},$ that the velocity of the flow to
the left of $x_{0}$ is $u$ and that the velocity of the flow to the right of 
$x_{1}$ is $\widetilde{u}.$ A nonviscous flow is a conservative system.
Equating the kinetic and potential energies of the flow for an arbitrary
surface particle of mass $m,$ we have 
\begin{equation*}
\frac{mu^{2}}{2}-\frac{m\widetilde{u}^{2}}{2}=mg\varepsilon. 
\end{equation*}
Writing the continuity equation in the form 
\begin{equation*}
\left( H+\delta\right) u=\left( H+\varepsilon\right) \widetilde{u}, 
\end{equation*}
the conservation of energy and continuity equations can be combined into the
single expression 
\begin{equation*}
\left( H^{2}+2H\varepsilon+\varepsilon^{2}\right) 2g\varepsilon=u^{2}\left(
2H+\varepsilon+\delta\right) \left( \varepsilon-\delta\right) 
\end{equation*}
or, to first order in $\varepsilon$ and $\delta,$%
\begin{equation*}
\varepsilon=\frac{\delta}{1-\frac{gH}{u^{2}}}=\frac{\delta}{1-\left( \frac
{c}{u}\right) ^{2}}. 
\end{equation*}

We find that the character of the flow is determined by whether the \textit{%
Froude number} $F=\sqrt{Q}/c$ exceeds, equals, or is exceeded by the number
1, where in this case of 1-dimensional flow, $\sqrt{Q}=\left| u\right| .$
The blow-up singularity that develops as $F$ tends to 1 \ is avoided by
hypothesis, as $\varepsilon$ is assumed to be small. In the case of \textit{%
tranquil flow} $(F<1),$ a positive elevation $\left| \varepsilon \right| $\
occurs when $\delta$ is exceeded by zero and a negative elevation $-\left|
\varepsilon\right| $\ occurs when $\delta$ exceeds zero. The opposite
relations hold for \textit{shooting flow} $(F>1).$ Note that this model
ignores the effects of turbulence.\newpage

\subsubsection{Technical note: The steady flow of shallow water}

\textit{``For not only the changes, but the tides themselves, are small with
respect to the magnitudes of the bodies in which they occur, though with
respect to us and our smallness they seem to be great things.''}

Galileo Galilei, \textsc{Dialogue Concerning the Two Chief World Systems}$%
^{21}$

\bigskip

Waves come with a natural scale determined by their wavelength. Considered
on this scale, virtually every other quantity associated with the tides is
small. In this note we derive some properties for water waves of small depth.

Write the flow velocity in components $\left( u,v,w\right) ,$ where $u$ is
the horizontal component in the $x$-direction, $v$ is the horizontal
component in the $y$-direction, and $w$ is the component in the $z$%
-direction. Impose initial conditions under which $w$ is zero at time $t=0.$
Applying the hydrostatic law, we conclude that $w=0$ for all subsequent
times and the horizontal velocity components $u$ and $v$ are independent of
the $z$-coordinate; thus we can generally adopt the notation of Sec. 3.1.
However, in this section $z$ represents the vertical direction and $(x,y)$
is a point on a horizontal cross section, \textit{e.g.,} the ocean surface.

Taking the flow to be steady, the velocity components are also independent
of $t$ and we can write the continuity equation in the form$^{22}$ 
\begin{equation*}
0=\frac{\partial}{\partial x}\left[ h\left( x,y\right) u\left( x,y\right) %
\right] +\frac{\partial}{\partial y}\left[ h\left( x,y\right) v\left(
x,y\right) \right] = 
\end{equation*}
\begin{equation}
u\frac{\partial h}{\partial x}+\frac{\partial u}{\partial x}h+v\frac{%
\partial h}{\partial y}+\frac{\partial v}{\partial y}h.
\end{equation}
We can use Bernoulli's formula to write $h$ as a function of $Q:$%
\begin{equation}
h\left( Q\right) =\frac{C-Q}{2g}.
\end{equation}
Substituting eq. (4) into eq. (3) and using the chain rule yields 
\begin{equation*}
\left[ \frac{C-Q}{2}-u^{2}\right] \frac{\partial u}{\partial x}-uv\left( 
\frac{\partial u}{\partial y}+\frac{\partial v}{\partial x}\right) +\left[ 
\frac{C-Q}{2}-v^{2}\right] \frac{\partial v}{\partial y}=0. 
\end{equation*}
The irrotationality of the flow implies that the velocity has vanishing
curl; in components, 
\begin{equation}
\frac{\partial u}{\partial y}-\frac{\partial v}{\partial x}=0.
\end{equation}
Writing 
\begin{equation*}
c^{2}=gh=\frac{C-Q}{2} 
\end{equation*}
and using (5), we obtain finally 
\begin{equation}
\left( c^{2}-u^{2}\right) \frac{\partial u}{\partial x}-2uv\frac{\partial u}{%
\partial y}+\left( c^{2}-v^{2}\right) \frac{\partial v}{\partial y}=0.
\end{equation}
Equations (5), (6) are formally analogous to the continuity equations for
the steady polytropic flow of an ideal gas in two dimensions.$^{23}$

As in the case of gas dynamics, the character of eq. (6) depends on whether
or not the flow speed $\sqrt{Q}$ exceeds the propagation speed $c.$ For 
\textit{subcritical} flow speeds in which the Froude number $F$ is exceeded
by 1, eq. (6) is of elliptic type. This equation type corresponds to closed
orbits in central force mechanics and to tranquil flow in hydraulics. In the
elliptic region only, the hodograph transformation taking (5), (6) to a
linear system will be nonsingular except at a finite number of points.$^{24}$
For \textit{supercritical} flow speeds in which $F$ exceeds 1, eq. (6) is of
hyperbolic type. This equation type corresponds to wave motion, most open
orbits, and shooting flow. In the hyperbolic region only, the system can be
attacked by the method of characteristics.$^{25}$ At the \textit{critical
value} $F=1,$ eq. (6) degenerates to a parabolic equation, the equation type
associated with diffusion. Shock waves are expected for values $F\geq1.^{26} 
$ Shocks which occur at the critical speed of a stationary gas flow have a
hydrodynamic analogue in hydraulic jumps, the stationary equivalent of a
tidal bore.$^{27}$

Equations (5), (6) can be derived by a variational principle from an energy
functional having the form 
\begin{equation*}
E=\int_{\Omega}\int_{0}^{Q}\widetilde{\rho}\left( s\right) dsd\Omega, 
\end{equation*}
where $\Omega$ is a surface and 
\begin{equation*}
\widetilde{\rho}\left( Q\right) =c^{2}. 
\end{equation*}
Precisely, it can be shown that eqs. (5), (6) are satisfied by differential
forms 
\begin{equation*}
d\varphi=udx+vdy 
\end{equation*}
which minimize the functional $E$ with respect to a class of competing
differential forms $d\psi$ having finite energy and for which the value $%
d\left( \varphi-\psi\right) $ integrates to a prescribed value on $%
\partial\Omega.$ This kind of variational approach goes back at least to the
work of Bateman on gas dynamics in the late 1920s.$^{28}$

Equation (5) implies the local existence of the potential $\varphi,$ which
is conventionally taken to be a (possibly multivalued) function in $\mathbb{R%
}^{2}.$ If, however, the potential is constrained to lie on a prescribed
surface, then eqs. (5), (6) transform into a generalized version of the
harmonic map equations in the subcritical region and of the wave map
equations in the supercritical region, and have wilder nonlinearities than
either harmonic maps or wave maps.$^{29}$

\section{Waves in a neighborhood of a caustic}

Imagine the sea surface to be a sheet of cloth, represented mathematically
by the $xy$-plane sitting in $\mathbb{R}^{3}.$ A calm sea might correspond
very roughly to a deformation of the cloth into a cylindrical surface in
which the values on the $z$ axis are given by a sine function, or
periodically extended gaussian function, of $x.$ Because such functions are
infinitely differentiable,$^{30}$ this surface can be projected by a smooth
coordinate transformation onto the plane $z=0.$ Considering various static
charges and air pockets that would be present in the material, we could not
expect the cloth to be absolutely flat as is the $xy$-plane; but any small
deformations are unstable in the sense that they could be removed by an
arbitrarily small rearrangement of the cloth. If we were to crumple the
cloth and then press down on it, so that it resembled the drapery in a
renaissance painting, we would find that only two types of deformations
could occur $-$ again ignoring variations that could be eliminated by an
infinitesimally small local deformation. These are folds and pleats. Note
that the vertex of a pleat is a cusp.

For example, the map given explicitly by the transformation 
\begin{equation*}
\left( x,y\right) \rightarrow\left( x,y^{2}\right) 
\end{equation*}
has a fold along the line $x=0,$ along which the half-plane containing
negative values of $y$ is folded by the map onto the half-plane containing
positive values of $y.$ On the other hand, the map given explicitly by the
transformation 
\begin{equation*}
\left( x,y\right) \rightarrow\left( x,xy-y^{3}\right) 
\end{equation*}
has a cusp at the origin, which is the vertex of a pleat constructed by
folding the $xy$-plane along the parabola in the first and second quadrants.$%
^{31}$

It can be shown that fold and cusp singularities are the only stable
singularities that could occur in otherwise smooth mappings of the plane to
itself.$^{32}$

In the rough analogy between the cloth and the sea surface, a fold
singularity in the cloth would correspond to a wave of abnormally high
intensity, spread over a relatively large region in comparison with a cusp
singularity, in which the intensity is even higher but the affected region
is much smaller. The existence of static charges in the cloth might
represent chromatic aberration due to the fact that waves are formed by a
combination of various wavelengths. This leads to the formation of little
caustics at various points of the wave. There is a difference in scale,
however, as static charges, under normal conditions, do not play a major
role in the morphology of a piece cloth lying on a table. On the other hand,
it is quite possible that chromatic aberration significantly affects the
formation of a wave.$^{33}$

\subsection{Lagrangian manifolds}

The preceding analysis suggests that, as an alternative to describing a
caustic as an envelope of a family of rays, a caustic can be described as a
singularity in a certain kind of map. In order to understand this alternate
description, we return to the consideration of caustics which are observed
in nature as bright regions of light.

Consider an \textit{isotropic} medium, by which we exclude crystalline media
in which forces tend to propagate along preferred directions (\textit{e.g.,}
fracture planes). We further suppose the medium to be \textit{nondissipative}%
, meaning that we ignore energy losses due to viscosity or other kinds of
friction. Assume that a scalar, monochromatic electromagnetic field is
propagating through this medium with frequency $\omega_{0},$ away from any
electric charges. We model the space-dependent part of the field by the 
\textit{Helmholtz equation} 
\begin{equation}
\Delta u+k^{2}\widetilde{n}^{2}u=0,
\end{equation}
where $\widetilde{n}$ is the refractive index of the medium and $k$ is the 
\textit{wave number}. Mathematically, the Helmholtz equation arises from
substituting the simplest formula for an oscillatory wave of amplitude $u(%
\mathbf{x})$ into the wave equation. Because $k^{-1}$ is proportional to
wavelength, the geometrical optics approximation is physically valid in the
region of large values for $k.$

In the region of geometrical optics, the values of $k$ dominate over all
other mathematically relevant parameters. In this region solutions to (7)
are usually written as an asymptotic expansion in powers of $k,$ having the
form$^{34}$ 
\begin{equation}
u\left( \mathbf{r}\right) \simeq\left( \sum_{j=0}^{\infty}U_{j}\left( 
\mathbf{r}\right) \cdot\left( ik\right) ^{-j}\right) \exp\left[ ik\psi\left( 
\mathbf{r}\right) \right] ,
\end{equation}
where $U_{j}$ and $\psi$ are scalar fields and $\mathbf{r}$ denotes the
radial vector in $\mathbb{R}^{3}.$ Substituting expansion (8) into eq. (7)
and equating coefficients of equal powers of $k,$ we obtain the system 
\begin{equation}
\left| \nabla\psi\right| ^{2}=\widetilde{n}^{2}(\mathbf{r}),
\end{equation}
and 
\begin{equation}
2\nabla\psi\cdot\nabla U_{0}+\left( \Delta\psi\right) U_{0}=0,
\end{equation}
\begin{equation}
2U_{0}\nabla\psi\cdot\nabla\left( \frac{U_{j}}{U_{0}}\right) +\Delta
U_{j-1}=0
\end{equation}
for $j=1,2,3,...$ . Equation (9) is the \textit{eikonal equation} and eqs.
(10), (11), the \textit{transport equations}.

Although the system (9)-(11) is a good deal more complicated than the system
(1), (2), the method of characteristics can be applied to both systems with
similar results. That is, we compute in this case as well the characteristic
curves (which represent light rays) and seek the envelope of that family of
curves.$^{35}$

It is conventional to write the equations for the characteristics of system
(9)-(11) in the canonical form 
\begin{equation*}
\frac{d\mathbf{r}}{d\tau}=\mathbf{p}, 
\end{equation*}
\begin{equation*}
\frac{d\mathbf{p}}{d\tau}=\frac{1}{2}\nabla\widetilde{n}^{2}. 
\end{equation*}
These are Hamilton's equations for position $\mathbf{r}=\left\{
x,y,z\right\} $ and momentum $\mathbf{p}=\left\{ p_{x},p_{y},p_{z}\right\} ,$
but ``momentum'' is not a physically obvious concept in the context of a
light wave. Because the vector $\nabla\psi$ lies in the direction in which
the wave propagates (the direction of a ray through the point $\mathbf{r}$),
we take 
\begin{equation*}
\mathbf{p}=\nabla\psi 
\end{equation*}
and choose $\tau$ so that 
\begin{equation*}
d\tau=\frac{ds}{\widetilde{n}}, 
\end{equation*}
where $s$ denotes arc length along the ray.

Thus if the field is initially localized on a surface $\Omega$ having
coordinates $\left( \xi,\eta\right) ,$ the ray consists of the trajectory 
\begin{equation*}
\mathbf{r}=\mathbf{R}\left( \xi,\eta,\tau\right) 
\end{equation*}
such that $\mathbf{R}\left( \xi,\eta,0\right) $ lies on $\Omega.$ Writing $%
\mathbf{\xi}=\left( \xi,\eta,\tau\right) ,$ we can consider either the 
\textit{ray surface} $\left\{ \mathbf{r},\mathbf{\xi}\right\} $ or the
6-dimensional phase space 
\begin{equation*}
\left\{ \mathbf{r},\mathbf{p}\right\} =\left\{
x,y,z;p_{x},p_{y},p_{z}\right\} 
\end{equation*}
\begin{equation*}
=\left\{ \mathbf{R}\left( \xi,\eta,\tau\right) ,\mathbf{p}\left( \xi
,\eta,\tau\right) \right\} . 
\end{equation*}
The latter is called a \textit{Lagrangian manifold}.$^{36}$

The projection map taking the ray surface onto the physical space having
local coordinates $\left\{ x,y,z\right\} $ is \textit{regular} at points for
which the Inverse Function Theorem holds. Points which are not regular, such
as those on folds or pleats, are \textit{singular} and are associated with a
vertical tangent plane. The tangent plane has the property that its
projection onto the $xy$-plane has dimension 2 except where it is vertical,
at which points it has dimension 1.

We can describe this property analytically by introducing the \textit{%
derivative map} $D\mathbf{f}$ from the plane to the surface, which is simply
the matrix of partial derivatives of the map\textbf{\ }$\mathbf{f}=\left(
f^{1},f^{2}\right) =\left( f^{1}\left( x,y\right) ,f^{2}\left( x,y\right)
\right) ,$ $\left( x,y\right) \in\mathbb{R}^{2}:$%
\begin{equation*}
D\mathbf{f}=\left[ 
\begin{array}{cc}
\partial f^{1}/\partial x & \partial f^{1}/\partial y \\ 
\partial f^{2}/\partial x & \partial f^{2}/\partial y
\end{array}
\right] . 
\end{equation*}
Singularities of the map $\mathbf{f},$ because they correspond to points at
which the tangent plane is vertical, correspond to points at which its
projection onto the $xy$-plane has dimension less than two, and thus to
points at which the matrix $D\mathbf{f}$ has less than maximal rank. (This
is, of course, equivalent to the vanishing of the Jacobian of the
transformation at such points, as the Jacobian is the determinant of the
matrix of derivatives.)

If, for example, we define 
\begin{equation*}
\mathbf{f}\left( x,y\right) =\left( \xi ,\eta \right) =\left(
x,xy-y^{3}\right) ,
\end{equation*}
then 
\begin{equation*}
D\mathbf{f}=\left[ 
\begin{array}{cc}
1 & 0 \\ 
y & x-3y^{2}
\end{array}
\right] .
\end{equation*}
This matrix has less than maximal rank along the parabola 
\begin{equation*}
x=3y^{2}.
\end{equation*}
All the points on this parabola correspond to fold singularities except the
origin of $\mathbb{R}^{2},$ which is a cusp.$^{37}$

More generally, we can consider mappings $\mathbf{f}$ between any parameter
space and the associated physical space. If we define a \textit{caustic} to
be a point of the mapping\ at which the matrix $D\mathbf{f}$ has less than
maximal rank, then this may appear to be a second, independent definition.
But $D\mathbf{f}$ has less than maximal rank precisely at those points for
which rays which were initially distinct, merge.$^{38}$ In this sense the
definition of a caustic as an envelope of a family of rays and that of a
caustic as a singularity in the projection map of a Lagrangian manifold onto
a subspace of $\mathbb{R}^{3}$ are equivalent descriptions of the same
phenomenon.

\subsection{The caustic boundary layer}

Because the infinite series in eq. (8), representing wave amplitude, is
proportional to the square root of the local density of rays, the asymptotic
approximation (8) has an amplitude singularity at a caustic. In particular,
the validity of such an approximation does not extend to any region that
lies beyond a caustic. This is analogous to having a model of the behavior
of an ocean wave before it breaks, but not after.

This failure motivates a more delicate approximation.$^{39}$ Consider the
mathematical problem of solving eq. (7) with refractive index $\widetilde
{n}\equiv1$ in the geometrical optics region of large $k.$ Replace eq. (8)
by an expansion having the form 
\begin{equation*}
u\left( x,y\right) = 
\end{equation*}
\begin{equation}
\left\{ g_{0}\left( x,y\right) V\left[ k^{2/3}\rho_{L}\left( x,y\right) %
\right] +\frac{g_{1}\left( x,y\right) }{ik^{1/3}}V^{\prime}\left[
k^{2/3}\rho_{L}\left( x,y\right) \right] \right\} \exp\left[ ik\theta\left(
x,y\right) \right] ,
\end{equation}
where $g_{0}\left( x,y\right) ,$ $g_{1}\left( x,y\right) ,$ $\rho _{L}\left(
x,y\right) ,$ and $\theta\left( x,y\right) $ are functions to be determined
with the solution; the function $V(t)$ is a solution of the \textit{Airy
equation} 
\begin{equation*}
V^{\prime\prime}\left( t\right) +tV\left( t\right) =0, 
\end{equation*}
originally introduced in Airy's model of the rainbow caustic.

On one side of the caustic, eq. (12) gives the oscillatory solution
predicted by geometrical optics. This is called the \textit{illuminated zone}
of the caustic. On the other side, called the \textit{shadow zone,} eq. (12)
gives an exponentially damped solution. The approximation remains bounded on
the caustic itself. That region can be interpreted as a boundary layer of
width proportional to $k^{-2/3},$ in which the solution undergoes a smooth
transition from the shadow zone to the illuminated zone.$^{40}$

In order to determine the four unknown functions it is necessary to
introduce four other equations, giving an expansion to second order in $k.$
The first two of the four auxiliary equations are equivalent to the eikonal
equation, and the last two are equivalent to the transport equations.

The terms to highest order in $k$ vanish if 
\begin{equation}
\left( \nabla\theta\right) ^{2}+\rho_{L}\left( \nabla\rho_{L}\right)
^{2}-1=0,
\end{equation}
\begin{equation}
2\nabla\theta\cdot\nabla\rho_{L}=0.
\end{equation}
The system (13), (14) is elliptic in the illuminated region for which $%
\rho_{L}$ exceeds zero but hyperbolic in the shadow region in which $\rho
_{L}$ is exceeded by zero. The critical value of the system (13), (14) is
defined by the caustic region in which $\rho_{L}$ vanishes.

In the region in which $\rho_{L}$ exceeds zero, we can multiply eq. (14) by $%
\pm\sqrt{\rho_{L}}$ and add eq. (13) to obtain in the illuminated zone the
eikonal equation (9) in variables 
\begin{equation*}
\psi^{\pm}\equiv\theta\pm\frac{2}{3}\rho^{2/3}. 
\end{equation*}

Terms of next highest order in $k$ vanish in the expansion provided 
\begin{equation*}
2\nabla\theta\cdot\nabla g_{0}+\Delta\theta
g_{0}+2\rho_{L}\nabla\rho_{L}\cdot\nabla g_{1} 
\end{equation*}
\begin{equation}
+\rho_{L}\Delta\rho_{L}g_{1}+\left( \nabla\rho_{L}\right) ^{2}g_{1}=0,
\end{equation}
\begin{equation}
2\nabla\rho_{L}\cdot\nabla
g_{0}+\Delta\rho_{L}g_{0}+2\nabla\theta\cdot\nabla g_{1}+\Delta\theta
g_{1}=0.
\end{equation}
In the illuminated zone we can employ the asymptotic expansions of the Airy
functions $V$ and $V^{\prime}$ to transform eqs. (15) and (16) into the
transport eqs. (10), (11) in variables 
\begin{equation*}
z^{\pm}=\frac{g_{0}\pm\sqrt{\rho_{L}}g_{1}}{\sqrt[4]{\rho_{L}}}. 
\end{equation*}
In comparison with the transport equations, the transformed equations (15),
(16) contain an extra term. This term has a physical interpretation as a
phase shift caused by the presence of the caustic as the wave passes from
shadow to light.

\subsubsection{Technical note: Singularities which are artifacts of the
linearization method}

Very recently,$^{41}$ boundary-value problems associated with the hodograph
linearization of eqs. (13), (14) have been introduced. The hodograph
transformation is generally singular. However, in the case of eqs. (5), (6)
it is known that the Jacobian of the hodograph transformation does not
vanish on any arc for nonconstant subcritical flow (see endnote 24). \ An
analogous assertion can be proven for eqs. (13), (14).

\bigskip

\textbf{Theorem}\textit{\ Let }$(u,v)$\textit{\ satisfy eqs. (13), (14).
Then the hodograph mapping } 
\begin{equation*}
\widetilde{h}:\left( u(x,y),v(x,y)\right) \rightarrow\left( \left(
x(u,v),y(u,v)\right) \right) 
\end{equation*}
\textit{and its inverse can only be singular at isolated points in the
elliptic region unless }$u$\textit{\ and }$v$\textit{\ are constants.}

\textit{\bigskip}

As the proof is very simple, we outline it:

We can write eqs. (13), (14) in the form$^{42}$%
\begin{equation}
\left[ f\left( u,v\right) -v^{2}\right] u_{x}+uv\left( v_{x}+u_{y}\right) +%
\left[ f\left( u,v\right) -u^{2}\right] v_{y}=0,
\end{equation}
\begin{equation}
u_{y}-v_{x}=0,
\end{equation}
for 
\begin{equation*}
f\left( u,v\right) \equiv\left( u^{2}+v^{2}\right) ^{2}. 
\end{equation*}
There is some interest in proving the theorem for general finite $f(u,v).$
For example, if $f\equiv1,$ then the hodograph image of (17), (18) is
identical to the Hodge equations on the projective disc.$^{43}$

The elliptic region of (17), (18) is defined by the inequality $%
u^{2}+v^{2}<f.$ \ The hodograph image of that system has the form 
\begin{equation}
\left( f-u^{2}\right) x_{u}-2uvx_{v}+\left( f-v^{2}\right) y_{v}=0,
\end{equation}
\begin{equation}
x_{v}=y_{u}.
\end{equation}
Using these equations, the Jacobian of the transformation to the hodograph
plane can be written 
\begin{equation*}
J=x_{u}y_{v}-x_{v}y_{u}=x_{u}y_{v}-x_{v}^{2} 
\end{equation*}
\begin{equation}
=\frac{\left[ \left( f-u^{2}\right) x_{u}-uvx_{v}\right] ^{2}+f\left(
f-u^{2}-v^{2}\right) x_{v}^{2}}{-\left( f-u^{2}\right) \left( f-v^{2}\right) 
}.
\end{equation}
Given that $f$ is nonnegative, we find that if $u^{2}+v^{2}<f,$ then $J$ is
negative unless $x_{u}=x_{v}=0.$ But this can only happen if $y_{u}$
vanishes as well, because of eq. (20). So the only possibly nonvanishing
partial derivative, when $J=0,$ is $y_{v}.$ But that possibility is
excluded, given the vanishing of $x_{u}$\ and $x_{v},$ by eq. (19). Now
Holmgren's Uniqueness Theorem$^{44}$ implies that in fact $x$ and $y$ must
be constants everywhere in the elliptic region if $J$ vanishes along any arc
in that region. Otherwise, $x$ and $y$ satisfy a linear, elliptic partial
differential equation having analytic coefficients, forcing $J$ to be
analytic in the elliptic region. The theory of analytic functions now
implies the assertion of the theorem. A similar argument can be constructed
for the Jacobian of the inverse transformation.

\bigskip

Substituting $f\left( u,v\right) =1$ into the extreme right-hand side of eq.
(21), we find that $J=0$ on the unit $(u,v)$-circle whenever the expression 
\begin{equation*}
\left( f-u^{2}\right) x_{u}-uvx_{v}=v\left( vx_{u}-ux_{v}\right) 
\end{equation*}
vanishes. This occurs on the $u$-axis regardless of the values of $x_{u}$
and $x_{v}.$ Thus the conclusion that $J$ does not vanish on an arc for
nonconstant $x_{u}$ and $x_{v}$ does not extend into the parabolic region of
the equations, which consists of points on the unit $\left( u,v\right) $%
-circle.

\subsection{Rogue waves}

\textit{``Similar to this and much greater, we understand, are the currents
between Africa and the very large island of Madagascar, as the waters of the
Indian and South Atlantic oceans which surround it flow and become
constricted in the smaller channel between it and the South African coast.''}

Galileo Galilei, to Alessandro Cardinal Orsini

\bigskip

The treacherous waters of the Agulhas, off the southeastern coast of South
Africa, have inspired many sea stories, and more than a few papers in
theoretical hydrodynamics.$^{45}$ The Agulhas Current is fed from the north
by the Mozambique and East Madagascar currents; most of its water, however,
is apparently derived from recirculation in the southwest Indian Ocean
subgyre. The Agulhas Current is very swift, reaching surface speeds of 2 $%
m/s.$ In addition the region is known for the spontaneous appearance of
solitary waves, \textit{rogue waves,} of enormous size that appear without
warning and are often unconnected with any unusual meteorological event.

Rogue waves are also hypothesized to result from the focusing of a wave at a
caustic; but in this case the focusing agent is not collision of the wave
with a channel entrance or a sloping beach, but rather collision of the wave
with a swift oncoming current. This induces a dramatic change in velocity
which, again, corresponds to the change in the velocity of light upon
encountering a change in the refractive index of the ambient medium.

When the dynamics of ocean waves are considered, the angle at which the wave
meets the incoming current appears to be significant. If the wave meets the
current head-on, the wave will be slowed more in the middle, where the
current is fastest, than at the ends. This could lead to a fold singularity;
but if the wave meets the current at an angle, shear effects might lead to a
more lethal cusp singularity (localized in a much smaller region).$^{46}$

In the interaction of a wave with an opposing current, a ray approximation
is indicated by comparison of the length of even large wind-induced swells,
such as the ones correlated with accidents in the Agulhas, with the scale of
horizontal variations of the current.$^{47}$

\section{Amphidromic points}

There are points of every ocean where there is no tide. These points are
singularities in the wavefront of the tides. But they are singularities in
the phase of the advancing wave rather than in its amplitude.

\subsection{Wavefronts}

\textit{``... whenever water is made to flow in this or that direction by a
noticeable retardation or acceleration of its containing vessel, it rises
here and subsides there, it does not however remain in such a state. Rather,
by virtue of its own weight and natural inclination to balance and level
itself out, it goes back with speed and seeks the equilibrium of its parts;
and, being heavy and fluid, not only does it move toward equilibrium but,
carried by its own impetus, it goes beyond ... This is similar to the way in
which a pendulum, after being displaced from the perpendicular,
spontaneously returns to it and to rest, but not before going beyond it many
times with a back and forth motion.''}

Galileo Galilei, \textit{op. cit.}

\bigskip

The wavefront of a light source can be imagined as a foliation of spheres,
each centered at the light source and expanding into space at the speed of
light in the medium. A two-dimensional version for water waves can be
observed by tossing a pebble into a still pond. The wavefront in that case
is a foliation of circles centered at the collision point and expanding at
the propagation speed for the medium, the constant $c$ of Sec. 3.1. These
expanding spheres and circles are examples of wavefronts. We can think of
wavefronts more generally as surfaces in $\mathbb{R}^{3},$ or curves in $%
\mathbb{R}^{2},$ which evolve in the direction of their normal vector.

In particular, the \textit{wavefront} $\mathbf{W}(t)$ of any smooth curve $%
M\subset\mathbb{R}^{2}$ at time $t>0$ can be defined as the set of vectors 
\begin{equation*}
\mathbf{W}_{\pm}\left( t\right) =\left\{ x\widehat{\mathbf{i}}+y\widehat{%
\mathbf{j}}\pm t\widehat{\mathbf{n}}\left( x,y\right) |\left( x,y\right) \in
M\right\} , 
\end{equation*}
where $\widehat{\mathbf{n}}\left( x,y\right) $ is the unit normal vector to $%
M$ at the point $\left( x,y\right) .$ \ If $M$ is the graph of a function $%
y=f(x)$, then 
\begin{equation*}
\widehat{\mathbf{n}}\left( x\right) =\frac{-f^{\prime}\left( x\right) }{%
\sqrt{1+\left[ f^{\prime}\left( x\right) \right] ^{2}}}\widehat {\mathbf{i}}+%
\frac{1}{\sqrt{1+\left[ f^{\prime}\left( x\right) \right] ^{2}}}\widehat{%
\mathbf{j}}. 
\end{equation*}
The two directions $\mathbf{W}_{+}$ and $\mathbf{W}_{-}$ correspond to the
two possible directions of the unit normal $\widehat{\mathbf{n}}$ at a point
on a plane curve.

A sphere or circle evolving in the direction of its outward-pointing normal
will be nonsingular; either will contract to a singularity at its center if
it evolves in the direction of its inward-pointing normal. Singularities of
other simple wavefronts can be easily computed. As an example we analyze the
evolution of the wavefront corresponding to a source having the geometry of
a quartic curve.$^{48}$

If $y=x^{4},$ then the wavefront propagating in the direction of the
outward-pointing normal will propagate along the negative $y$-axis. \ The
wavefront in this case will be given by 
\begin{equation*}
\mathbf{W}_{-}\left( t\right) =\left( x+\frac{4x^{3}t}{\sqrt{1+\left(
4x^{3}\right) ^{2}}}\right) \widehat{\mathbf{i}}+\left( x^{4}-\frac {t}{%
\sqrt{1+\left( 4x^{3}\right) ^{2}}}\right) \widehat{\mathbf{j}}. 
\end{equation*}
This wavefront is of course a curve. \ By definition, singularities of this
curve are points at which the components of the tangent vector vanish. The
tangent vector 
\begin{equation*}
\mathbf{T}_{-}(x)\equiv\sigma_{1_{-}}^{\prime}\left( x\right) \widehat {%
\mathbf{i}}+\sigma_{2_{-}}^{\prime}\left( x\right) \widehat{\mathbf{j}} 
\end{equation*}
corresponding to the curve 
\begin{equation*}
\mathbf{W}_{-}\left( t\right) =\sigma_{1_{-}}\left( x\right) \widehat{%
\mathbf{i}}+\sigma_{2_{-}}\left( x\right) \widehat{\mathbf{j}} 
\end{equation*}
will vanish at values of $t$ for which 
\begin{equation*}
\sigma_{1_{-}}^{\prime}\left( x\right) =\sigma_{2_{-}}^{\prime}\left(
x\right) =0. 
\end{equation*}
But in our case, 
\begin{equation*}
\sigma_{1_{-}}^{\prime}\left( x\right) =1+\frac{12x^{2}t}{\left[ 1+\left(
4x^{3}\right) ^{2}\right] ^{3/2}}, 
\end{equation*}
which vanishes for no positive value of $t$ (so it never vanishes). \ Thus a
wavefront propagating along the outward-pointing normal vector will be
nonsingular for all time.

On the other hand, the wavefront moving along the inward-pointing normal to
the source curve is given by 
\begin{equation*}
\mathbf{W}_{+}\left( t\right) =\left( x-\frac{4x^{3}t}{\sqrt{1+\left(
4x^{3}\right) ^{2}}}\right) \widehat{\mathbf{i}}+\left( x^{4}+\frac {t}{%
\sqrt{1+\left( 4x^{3}\right) ^{2}}}\right) \widehat{\mathbf{j}}. 
\end{equation*}
In this case 
\begin{equation*}
\sigma_{1_{+}}^{\prime}\left( x\right) =1-\frac{12x^{2}t}{\left[ 1+\left(
4x^{3}\right) ^{2}\right] ^{3/2}} 
\end{equation*}
and 
\begin{equation*}
\sigma_{2_{+}}^{\prime}\left( x\right) =4x^{3}-\frac{48x^{5}t}{\left[
1+\left( 4x^{3}\right) ^{2}\right] ^{3/2}}. 
\end{equation*}
The wavefront in this direction will develop singularities at time 
\begin{equation*}
t=\frac{\left[ 1+\left( 4x^{3}\right) ^{2}\right] ^{3/2}}{12x^{2}}. 
\end{equation*}

The surface waves of the tides also possess wavefronts. But because the
source of the wave is a periodic force exerted on the water rather than the
pulse created by the collision of a pebble with the water, the wavefront of
the tides expands and contracts. In the absence of displacements due to the
land, to viscosity and friction with the ocean basin, and to currents pushed
by prevailing winds such as the Gulf Stream, the waves of the tides would
expand and contract in harmonic motion like a simple pendulum subjected to a
gentle disturbance.

If the tides were standing waves and their oscillations were only vertical,
then we could write them in the form of a real-valued function 
\begin{equation*}
\psi_{s}=\left( 2a\cos\omega t\right) \sin kx=\alpha\left( t\right) \sin kx, 
\end{equation*}
where $a$ is the amplitude of each of the two individual waves, $\omega$ is
the angular frequency and $k$ is the wave number. Nodal points of the
standing wave occur whenever $x$ is an integral multiple of $\pi/k.$ Zeros
of $\alpha\left( t\right) $ are not isolated points of the wave, but lines
along which the displacement of the wave vanishes for every value of $x.$

But the net force exerted on the tides is mainly horizontal$^{49}$: the high
tide moves in a \textit{tidal current} across the ocean and up onto the
land. Thus a standing-wave representation of the form $\psi_{s},$ for which
the zeros of $\alpha$ are nodal lines of a vertically oscillating wave, is
not a realistic model of the tides.$^{50}$

In exact analogy with electromagnetism, a description of the tidal current
requires a representation in terms of complex vector fields. In fact, the
elliptical orbit \textit{(polarization ellipse)} of the electric field
vector in a paraxially propagating electromagnetic field has an exact
analogy in the elliptical orbit of the horizontal velocity vector of a tidal
current. This implies the possibility of \textit{polarization singularities,}
at which the velocity ellipses degenerate to circles, or the related class
of singularities manifested in \textit{polarization lines,} describing
linear oscillations. Velocity ellipses of the tidal currents were studied by
W. Hansen in the early 1950s; but the interpretation of their degeneracies
as polarization singularities, with the explicit optical analogy, is recent.$%
^{51}$ We will say no more about them, proceeding to the simpler
singularities of phase which are associated to the complex representation of
scalar waves given in Sec. 2. Scalar waves cannot, of course, have
polarization singularities, but they can and do possess wavefront
dislocations.

Apply the complex scalar representation of Sec. 2.2 to the vertical
component of surface waves created by tidal forces. The wavefronts, curves
of constant phase, will correspond to \textit{cotidal lines.} These are
curves drawn through adjacent points of the ocean having high tide at the
same time. Executing such a survey for the North Sea, Whewell found two
points at which all the neighboring cotidal lines intersected. Such points
would be permanently at high tide. Whewell inferred two rotary systems of
waves centered at points at which the tide is high at all hours.$^{52}$
Although these \textit{amphidromic points} are located on the surface of
nautical charts, they should be conceived as projections onto the surface of
a singular line, a wavefront dislocation of the kind discussed in Sec. 2.$%
^{53}$ The vertical oscillations at such points have zero amplitude, leaving
the phase undefined.

That amphidromic points are true phase singularities of the tidal flow is
indicated by the fact that the tidal current about the associated
dislocation line does not integrate to zero, but behaves analogously to the
electric current surrounding a point charge or a current of water in the
presence of a source or sink. That is, define the current 
\begin{equation*}
\mathbf{j}=\alpha^{2}\nabla\chi, 
\end{equation*}
where $\alpha$ is the wave amplitude and $\chi$ is its phase, and imagine
that the current flow executes a simple closed loop $\gamma$ about the
dislocation line. Then the integral of the phase change about the loop
satisfies 
\begin{equation*}
\oint_{\gamma}d\chi=\pm2\pi 
\end{equation*}
no matter how the circuit is varied, provided that $\gamma$ remains
topologically equivalent to a circle about the dislocation line. It is for
this reason that wave dislocations in optics, which have an analogous
property, are sometimes called \textit{optical vortices.} 
\begin{equation*}
\ast\;\;\;\ast\;\;\;\ast 
\end{equation*}

\textit{``I could propose many other considerations if I wanted to delve
into finer details. Many, many more could be advanced if we had abundant,
clear, and truthful empirical reports of observations ... At the moment I
only claim to have given something of a sketch ... I hope, however, that it
does not turn out to be delusive, like a dream which gives a brief image of
truth followed by an immediate certainty of falsity.''}

\bigskip

\textbf{Notes}

\bigskip

$^{1}$Translation: [Finocchiaro, 1989]. This \textit{Discourse on the Tides}
was later expanded by Galileo into the Fourth Day of his \textit{Dialogue
Concerning the Two Chief World Systems} [Drake, 1967].

$^{2}$However, the resulting tide-raising force varies as a cube of the
distance to the celestial body; see, \textit{e.g.}, [Tricker, 1965],
Appendix I or [Elmore \& Heald, 1985], Sec. 6.4b.

$^{3}$Newton directly addressed the origin of tides in Book I, Proposition
66, and Book III, Proposition 24, of the \textit{Principia}. This paragraph
is a crude simplification of those arguments. Note that vertical
oscillations of the tides can be represented by a scalar field, whereas
their horizontal oscillations require a vector field for their description; 
\textit{c.f.} Sec. 5.1.

$^{4}$[Finocchiaro, 1989], Ch. IV, Sec. 4, pp. 127-131.

$^{5}$However, an early illustration of hydraulic jumps, which are
mathematically analogous to bores but are stationary rather than
progressive, can be found in the notebooks of Leonardo da Vinci, and Giorgio
Bidone described a hydraulic jump in 1820. Recurring tidal bores on the
English rivers Severn and Trent, the French river Seine near
Caudebec-en-Caux, and the Chinese river Tsien-Tang have been well known for
many years. See [Tricker, 1965], Ch. V, [Bascom, 1980], p. 104, and
[Chanson, 2001] for discussions. This last reference lists other examples of
bores of various sizes which have been observed in Mozambique, Malaysia,
Brazil, Australia, Canada, Alaska, and France.

$^{6}$See page 462 of [Drake, 1967], in which Galileo ridicules Kepler for
his accurate kinematic theory of the tides \textit{(``I am more astonished
at Kepler than at any other...'')}. There is a Hungarian proverb, ``It is
not enough to be wrong, one must also be rude.''

$^{7}$The scientific content of Galileo's theory of the tides is debated in
a number of references cited in [Finocchiaro, 1989], note 3 to Chapter IV.

$^{8}$See the collection [Soskin, 1998].

$^{9}$Historical remarks can be found in [Berry, 2000] and [Berry, 2002].

$^{10}$An early illustration of a nephroidal catacaustic, with a graphical
solution by ray tracing, can be found in the notebooks of Leonardo da Vinci.

$^{11}$See, \textit{e.g.,} [Berry \& Upstill, 1980] for a review.

$^{12}$It is observed, however, in [Born \& Wolf, 1999] that the first
reference to diffraction appears in the work of Leonardo da Vinci.

$^{13}$See, \textit{e.g.,} [Berry, 1998].

$^{14}$These approximations are reviewed in Secs. 2.1 and 2.2 of [Stoker,
1957]. See in particular, pp. 24, 25. See also Secs. 6.2 and 6.4\textit{a})
of [Elmore \& Heald, 1985].

$^{15}$However, results have been proven for various special circumstances.
See [Nalimov, 1974], [Kano \& Nishida, 1979], [Yoshihara, 1982], [Craig,
1985], [Wu, 1997], and [Schneider \& Wayne, 2000]. For a recent review of
the problem in the context of Korteweg-de Vries and nonlinear
Schr\"{o}dinger models (which we do not discuss), see [Schneider \& Wayne,
2002].

$^{16}$Details are given in Secs. 2 and 3 of [Stoker, 1947], which we
generally follow for most of Sec. 3.1. This reference corresponds to Secs.
10.1-10.5 of [Stoker, 1957].

$^{17}$This is done in Secs. 7 and 8 of [Stoker, 1947], which corresponds to
Secs. 10.6 and 10.7 of [Stoker, 1957]. Stoker's approach has become, by this
point in time, rather classical. It is generally preferable to formulate
problems for singular solutions to differential equations in the form of 
\textit{weak solutions,} an extension of the notion of a solution to rougher
function spaces than the spaces of multiply differentiable functions.
Because (1) and (2) are essentially Burgers equations, the discussion in
[Smoller, 1983], Ch. 15, of weak solutions with jump discontinuities applies
to them. See also the discussion of weak solutions in [Ladyzhenskaya \&
Ural'tseva, 1968].

$^{18}$Details are given in [Carrier \& Greenspan, 1958]. See also Sec.
2.6.2 of [Johnson, 1997].

$^{19}$See [Tricker, 1965], page 61.

$^{20}$Details of the model that we are about to describe can be found in
Ch. V of [Tricker, 1965], in which reference one can also find analogous
computations for the case of a decrease in the horizontal width of the
channel, as well as other extensions.

$^{21}$Translation: [Drake, 1967].

$^{22}$Equation (3) is derived in Sec. 2.6 of [Johnson, 1997] and Sec. 1.1.2
of [Mei, 1989].

$^{23}$Equations (5) and (6) are eqs. (10.12.2) and (10.12.5) of [Stoker,
1957]. The time-dependent form of these equations is derived in (A1)-(A3) of
the appendix to [Guza \& Bowman, 1976], where they are written in terms of
the velocity potential $\varphi.$ Regarding the analogy to gas dynamics,
compare eqs. (5) and (6) with eq. (14) of [Bers, 1958], taking $%
u=\varphi_{x},$ $v=\varphi_{y}.$ This analogy seems to have originated in
[Riabouchinsky, 1932].

$^{24}$For a proof, see, \textit{e.g.,} [Chapman, 2000], Sec. 12.4, or Sec.
14.3 of [Garabedian, 1998].

$^{25}$This is described in Ch. IV of [Courant \& Friedrichs, 1948].

$^{26}$See [Courant \& Friedrichs, 1948], especially, Secs. 30, 66, and 105;
also [Morawetz, 1982]. The existence and stability of transonic shocks in
solutions to the continuity equations for steady potential flow is
rigorously proven in [Chen \& Feldman, 2003].

$^{27}$This is described on pages 407, 408 of [Stoker, 1957] and in Sec. 2.7
of [Johnson, 1997].

$^{28}$See [Bateman, 1929]. The existence of subcritical flow was proven by
rigorous variational arguments in [Shiffman, 1952]. See also [Dong \& Ou,
1993].

$^{29}$Subcritical potential flow with geometric constraints is studied in
[Otway, 2000] and [Otway, 2004a].

$^{30}$A function accurately representing the sea surface at reasonable
scales would not be differentiable at all. Not only does the sea surface
fail to have a gaussian cross section, the statistical distribution of
typical wave \textit{amplitudes} is not even gaussian; see [Feder, 1988],
Ch. 11. A sine-wave approximation of water waves is realistic only for small
amplitudes. It has become quite conventional, by the way, to illustrate the
classification of singularities of mappings from $\mathbb{R}^{2}$ to $%
\mathbb{R}^{2}$ with crumpled tablecloths, sandwich bags, and strips of
cellophane. Some artifacts of the model are noted in Sec. 2.6 of [Nye, 1999].

$^{31}$These and other examples are analyzed in [Callahan, 1974].

$^{32}$This is proven in [Whitney, 1955].

$^{33}$This suggests that the conventional frequency distribution of wave
size predicted by the Rayleigh distribution undercounts the frequency of
extreme water waves; see [Dean, 1990].

$^{34}$See [Lewis \& Keller, 1964] and [Duistermaat, 1978] for the
geometrical optics case. See [Keller, 1958] and the seond section of [Chao,
1971] for the case of water waves.

$^{35}$Details of the ensuing discussion can be found in Chs. 1-3 of
[Kravtsov \& Orlov, 1999]. See also Ch. 3 of [Born \& Wolf, 1999]. A
derivation of the eikonal equation for water waves, and its relation to
Fermat's principle, is given in Sec. 3.2 of [Mie, 1989]. For a mathematical
discussion of the eikonal equation, see Sec. II.6.1 of [Courant \& Hilbert,
1962].

$^{36}$In technical language, a Lagrangian manifold of a dynamical system is
a submanifold of the phase space, having dimension equivalent to the
dimension of the configuration space, on which the 2-form defining a
symplectic structure on the phase space vanishes identically. This 2-form is
the wedge product $dp\wedge dq,$ where the 0-form $p$ gives coordinates on
the momentum space and the 0-form $q$ gives coordinates on the configuration
space. See, \textit{e.g.,} Appendices 11 and 12 of [Arnold, 1989] and the
references therein. In that work the Helmholtz equation is replaced by the
Schr\"{o}dinger equation as a motivating example. (See in this respect Sec.
12.5 of [Poston \& Stewart, 1978].) This is important because in studies of
the effects of caustics on water waves, various equations have been used,
including a nonlinear Schr\"{o}dinger equation. Compare for example, [Chao,
1971], Sec. 3.2 of [Mie, 1989], and [Smith, 1976].

$^{37}$See Fig. 10 of [Callahan, 1974], Fig. 3.2 of [Kravtsov \& Orlov,
1999], or Sec. 2 of [Britt, 1985].

$^{38}$See, \textit{e.g.,} Sec. 2 of [White \& Fornberg, 1998] or Sec. 4 of
[Peregrine \& Smith, 1979], where this property is derived for water waves
under different hypotheses. Section 3.1 of [Berry \& Upstill, 1980] is an
exposition of this property in the context of catastrophe optics.

$^{39}$See [Kravtsov, 1964] and [Ludwig, 1966] for the geometrical optics
case, and the third section of [Chao, 1971] for the case of water waves. An
early review is [Berry, 1969].

$^{40}$This is Ludwig's interpretation.

$^{41}$[Magnanini \& Talenti, 2002].

$^{42}$\textit{c.f.} eq. (3.1) of [Magnanini \&\ Talenti, 2002], taking the
function $v(x,y)$ of that work to be the scalar potential for the vector $%
(u,v)$ of our eqs. (17), (18).

$^{43}$See Sec. 2 of [Otway, 2002]. Details of the following proof, in a
form applicable to either system, are given in [Otway, 2004b].

$^{44}$See, \textit{e.g.,} Appendix 2 to Chapter III of [Courant \& Hilbert,
1962] or pp. 185-188 of [Garabedian, 1998].

$^{45}$The former include the disappearance of the ``unsinkable'' ship 
\textit{Waratah,} possessing eight watertight compartments, in a storm which
she should have easily weathered; the sinking of the Greek cargo steamer 
\textit{Margarita} without a trace in 1925, after signalling a $20^{\circ}$
list; the report, also in 1925, by a South African air force pilot of a
submerged wreck off the Transkei coast in clear seas; the huge wave which
struck the steamer \textit{Rabaul} in 1936; and the massive ocean cavity
encountered by the 28,000-ton \textit{Edinburgh Castle} in 1964, causing a $%
30^{\circ}$ list. When Gamal Abdal Nasser closed the Suez Canal in 1967, oil
tankers returning from the Persian Gulf were forced to travel along the
southeastern coast of Africa, where the Agulhas current is encountered and
its speed can be exploited. By the time that the Suez Canal reopened, it had
become the practice to build tankers on a much larger scale than before,
making the narrow canal an unattractive alternative to rounding the Cape of
Good Hope. This has led to a number of tanker accidents, apparently due to
collision of the massive ships with rogue waves. These include the wave that
broke the tanker \textit{World Glory} in two in 1968; the wave that caused
the \textit{Neptune Sapphire} to lose part of its bow in 1973; the 1985 loss
of the Soviet tanker \textit{Taganrogsky Zaliv}; the 1986 incident involving
the semisubmersible drilling rig \textit{Actinia;} and the sinking of the
Greek passenger liner \textit{Oceanos} in 1991. The data had already begun
to be generalized into a coherent hydrodynamic theory by 1974, when a
professor at the University of Cape Town analyzed eleven accidents and
proposed that geological conditions in a narrow region between Richards Bay
and Cape Agulhas contribute to the production of rogue waves along the
Agulhas Current [Mallory, 1974]. Subsequently a number of mathematical
models for various aspects of this phenomenon have been proposed, including
[Peregrine, 1976], [Smith, 1976], [Gerber, 1996], [Lavrenov, 1998], [White
\& Fornberg, 1998], and [Pelinovsky \& Kharif, 2000]. A source for the
physical oceanography of Southeast African currents is [Stramma \&
Lutjeharms, 1997].

$^{46}$Of course this kind of classification is highly speculative, if for
no other reason than the obvious one, that ocean waves are not smooth
mappings. Some plausible diagrams for extreme waves formed by collision with
an oncoming current are given in [Poston \& Stewart, 1978], Figs. 12.45,
12.46. See also the review [Meyer, 1979] for arguments based on classical
analysis.

$^{47}$This point is made in [Smith, 1976], p. 417.

$^{48}$For a discussion of other examples, see [Barreto, 1997].

$^{49}$An elementary explanation of this property is given in Ch. 1 of
[Tricker, 1965]. A slightly more technical derivation is given in Sec.
6.4(d) of [Elmore \& Heald, 1985].

$^{50}$This observation is made in [Berry, 2001], Sec. 4, in which the
horizontal progression of the tides is connected to the breaking of
time-reversal symmetry by the axial rotation of the Earth.

$^{51}$See, \textit{e.g.}, Sec. 12.9 of [Nye, 1999] or Sec. 5 of [Berry,
2001]. For historical remarks, see [Berry, 2000].

$^{52}$The periodicities of the tides are many and complicated, due to the
combined effects of the gravitational attraction of the Sun and Moon.
Whewell measured the dominant $M_{2}$ component, having period 12 hours and
25 minutes.

$^{53}$See [Berry, 1998] for a discussion of how these lines are related to
singular points.

\bigskip

{\large References}

\bigskip

[Arnold, 1989] V. I. Arnold, \textit{Mathematical Methods of Classical
Mechanics,} Springer-Verlag, New York, 1989.

[Barreto, 1997] A. S\'{a} Barreto, Wave propagation and the formation of
caustics, \textit{Simmer Notes}, Fields Institute e-print, 1997.

[Bascom, 1980] W. Bascom, \textit{Waves and Beaches}, Anchor Books, Garden
City, 1980.

[Bateman, 1929] H. Bateman, Notes on a differential equation which occurs in
the two-dimensional motion of a compressible fluid and the associated
variational problem, \textit{Proc. R. Soc. London Ser. A,} \textbf{125}
(1929), 598--618.

[Berry, 1969] M. V. Berry, Uniform approximation: a new concept in wave
theory, \textit{Sci. Progr., Oxf.} \textbf{57} (1969), 43-64.

[Berry, 1998] M. V. Berry, Much ado about nothing: optical dislocation lines
(phase singularities, zeros, vortices...), in [Soskin, 1998].

[Berry, 2000] M. V. Berry, Making waves in physics. Three wave singularities
from the miraculous 1830s', \textit{Nature} \textbf{403}, 21-26, January,
2000.

[Berry, 2001] M. V. Berry, Geometry of phase and polarization singularities,
illustrated by edge diffraction and the tides, in: \textit{Second
International Conference on Singular Optics (Optical Vortices): Fundamentals
and applications, SPIE 4403} (Bellingham Washington), pp. 1-12.

[Berry, 2002] M. V. Berry, Exuberant interference: rainbows, tides, edges,
(de)coherence..., \textit{Phil. Trans. R. Soc. London A} (2002) \textbf{360}%
, 1023-1037.

[Berry \& Upstill, 1980] M. V. Berry and C. Upstill, Catastrophe optics:
Morphologies of caustics and their diffraction patterns, \textit{Progress in
Optics} \textbf{18} (1980), 257-346.

[Bers, 1958] L. Bers, \textit{Mathematical Aspects of Subsonic and Transonic
Gas Dynamics,} Wiley, New York, 1958.

[Born \& Wolf, 1999] M. Born and E. Wolf, \textit{Principles of Optics,}
Cambridge University Press, Cambridge, 1999.

[Britt, 1985] J. Britt, The anatomy of low dimensional stable singularities, 
\textit{Amer. Math. Monthly} \textbf{92} (1985), 183-201.

[Callahan, 1974] J. Callahan, Singularities and plane maps, \textit{Amer.
Math. Monthly} \textbf{81} (1974), 211-240.

[Carrier \& Greenspan, 1958] G. F. Carrier and H. P. Greenspan, Water waves
of finite amplitude on a sloping beach, \textit{J. Fluid Mech.} \textbf{4}
(1958), 97-109.

[Chanson, 2001] H. Chanson, Flow field in a tidal bore : a physical model,
in: \textit{Proc. 29th IAHR Congress, Beijing, China,} G. Li, ed., Tsinghua
University Press, Beijing, 2001, pp. 365-373.

[Chao, 1971] Y-Y. Chao, An asymptotic evaluation of the wave field near a
smooth caustic, \textit{J. Geophys. Res.} \textbf{76} (1971), 7401-7408.

[Chapman, 2000] C. J. Chapman, \textit{High Speed Flow,} Cambridge
University Press, Cambridge, 2000.

[Chen \& Feldman, 2003], C-Q Chen and M. Feldman, Multidimensional transonic
shocks and free boundary problems for nonlinear equations of mixed type, 
\textit{J. Amer. Math. Soc.} \textbf{16} (2003), 461-464.

[Courant \& Friedrichs, 1948] R. Courant and K. O. Friedrichs, \textit{%
Supersonic Flow and Shock Waves,} Interscience, New York, 1948.

[Courant \& Hilbert, 1962] R. Courant and D. Hilbert, \textit{Methods of
Mathematical Physics, Vol. II: Partial Differential Equations,}
Interscience, New York, 1962.

[Craig, 1985] W. Craig, An existence theorem for water waves and the
Boussinesq and Korteweg-de Vries scaling limits, \textit{Commun. Partial
Differential Equations} \textbf{10} (1985), 787-1003.

[Dean, 1990] R. G. Dean, Freak waves: a possible explanation, in: \textit{%
Water Wave Kinematics} (A. Torum and O. T. Gudmestad, eds.), Kluwer,
Dordrecht, 1990, pp. 609-612.

[Dong \& Ou, 1993] G. Dong and B. Ou, Subsonic flows around a body in space, 
\textit{Commun. Partial Differential Equations} \textbf{18} (1993), 355-379.

[Drake, 1967] G. Galilei, \textit{Dialogue Concerning the Two Chief World
Systems,} translated with revised notes by Stillman Drake, University of
California Press, Berkeley, 1967.

[Duistermaat, 1978] J. J. Duistermaat, The light in a neighborhood of a
caustic, in: \textit{Seminaire Bourbaki 1976/77,} Expos\'{e} no. 490,
Springer-Verlag, New York, 1978.

[Elmore \& Heald, 1985] W. C. Elmore and M. A. Heald, \textit{Physics of
Waves}, Dover, New York, 1985.

[Feder, 1988] J. Feder, \textit{Fractals}, Plenum, New York, 1988.

[Finocchiaro, 1989] M. A. Finocchiaro, \textit{The Galileo Affair},
University of California Press, Berkeley, 1989.

[Garabedian, 1998] P. Garabedian, \textit{Partial Differential Equations},
AMS, Providence, 1998.

[Gerber, 1996] M. Gerber, Giant waves and the Agulhas Current, \textit{%
Deep-Sea Res., 1996.}

[Guza \& Bowen, 1976] R. T. Guza and A. J. Bowen, Finite amplitude edge
waves, \textit{J. Marine Res}. \textbf{34}, No. 2 (1976), 269-293.

[Johnson, 1997] R. S. Johnson, \textit{A Modern Introduction to the
Mathematical Theory of Water Waves,} Cambridge, 1997.

[Kano \& Nishida, 1979] T. Kano and T. Nishida, Sur les ondes de surface de
l'eau avec une justification mathematique des equations des ondes en eau
profonde, \textit{J. Math. Kyoto Univ.} \textbf{19} (1979), 335-370.

[Keller, 1958] J. B. Keller, Surface waves on water of nonuniform depth, 
\textit{J. Fluid Mech.} \textbf{4} (1958), 607-614.

[Kravtsov, 1964] Yu. A. Kravtsov, A modification of the geometrical optics
method [in Russian], \textit{Radiofizika} \textbf{7} (1964), 664-673.

[Kravtsov \& Orlov, 1999] Yu. A. Kravtsov and Yu. I. Orlov, \textit{%
Caustics, Catastrophes, and Wave Fields}, Springer-Verlag, New York, 1999.

[Ladyzhenskaya \& Ural'tseva, 1968] O. A. Ladyzhenskaya and N. N.
Ural'tseva, \textit{Linear and Quasilinear Elliptic Equations,} Academic
Press, New York, 1968.

[Lavrenov, 1998] I. Lavrenov, The wave energy concentration at the Agulhas
current off South Africa, \textit{Natural Hazards} \textbf{17} (1998),
117-127.

[Lewis \& Keller, 1964] R. M. Lewis and J. B. Keller, Asymptotic methods for
partial differential equations: the reduced wave equation and Maxwell's
equation. Courant Institute, 1964.

[Ludwig, 1966] D. Ludwig, Uniform asymptotic expansions at a caustic, 
\textit{Commun. Pure Appl Math.} \textbf{19}, No. 2 (1966), 215-250.

[Magnanini \& Talenti, 2002] R. Magnanini and G. Talenti, Approaching a
partial differential equation of mixed elliptic-hyperbolic type, in \textit{%
Ill-posed and Inverse Problems}, S. I. Kabanikhin and V. G. Romanov, eds.,
VSP 2002, pp. 263-276.

[Mallory, 1974] J. K. Mallory, Abnormal waves on the south east coast of
South Africa, \textit{Intl. Hydrog. Rev.} \textbf{51} (1974), 99-129.

[Meyer, 1979] R. E. Meyer, Theory of water-wave refraction, \textit{Adv.
Applied Mech.} \textbf{19} (1979), 53-141.

[Mie, 1989] C. C. Mei, \textit{The Applied Dynamics of Ocean Surface Waves,}
World Scientific, Singapore, 1989.

[Morawetz, 1982] C. S. Morawetz, The mathematical approach to the sonic
barrier, \textit{Bull. Amer. Math. Soc. (N.S.)} \ \textbf{6} (1982), 127-145.

[Nalimov, 1974] V. I. Nalimov, The Cauchy-Poisson problem [in Russian], 
\textit{Dynamika Sploshnoi Sredy} \textbf{18} (1974), 104-210.

[Nye, 1999], J. F. Nye, \textit{Natural Focusing and the Fine Structure of
Light,} Institute of Physics Publishing, Bristol, 1999.

[Otway, 2000] T. H. Otway, Nonlinear Hodge maps, \textit{J. Math. Phys.} 
\textbf{41}, No. 8 (2000), 5745-5766. A slightly revised and updated version
is posted at arXiv:math-ph/9908030.

[Otway, 2002] T. H. Otway, Hodge equations with change of type, \textit{%
Annali di Matematica, Pura ed Applicata} \textbf{181} (2002), 437-452.

[Otway, 2004a] T. H. Otway, Maps and fields with compressible density, 
\textit{Rendiconti del Seminario Matematico dell' Universit\`{a} di Padova} 
\textbf{111} (2004), 133-159.

[Otway, 2004b] T. H. Otway, Geometric analysis near and across a sonic
curve, in: \textit{New Developments in Mathematical Physics Research,}
Charles V. Benton, ed., Nova Science Publishers, New York, 2004, pp. 27-54.

[Pelinovsky \& Kharif, 2000] E. Pelinovsky and C. Kharif, Simplified model
of the freak wave formation from the random wave field, \textit{Rogue Waves
2000,} Brest, 2000.

[Peregrine, 1976] D. H. Peregrine, Interaction of water waves and currents, 
\textit{Adv. Appl. Mech.} \textbf{16} (1976), 9-117.

[Peregrine \& Smith, 1979] D. H. Peregrine and R. Smith, Nonlinear effects
upon waves near caustics, \textit{Philos. Trans. R. Soc. London, Ser. A} 
\textbf{292} (1979), 341-370.

[Poston \& Stewart, 1978] T. Poston and I. Stewart, \textit{Catastrophe
Theory}, Pitman, London, 1978.

[Riabouchinsky, 1932] D. Riabouchinsky, Sur l'analogie hydraulique des
mouvements d'un fluide compressible, \textit{Comptes Rendus Academie des
Sciences, Paris} \textbf{195} (1932), 998.

[Schneider \& Wayne, 2000] G. Schneider and C. E. Wayne, The long wave limit
for the water wave problem. I. The case of zero surface tension, \textit{%
Commun. Pure Appl. Math.} \textbf{53}, No. 12 (2000), 1475-1535.

[Schneider \& Wayne, 2002] G. Schneider and C. E. Wayne, On the validity of
2D-surface water wave models, e-print.

[Shiffman, 1952] M. Shiffman, On the existence of subsonic flows of a
compressible fluid, \textit{J. Rat. Mech. Anal.} \textbf{1} (1952), 605-652.

[Smith, 1976] R. Smith, Giant waves, \textit{J. Fluid Mech.} \textbf{77}
(1976), 417-431.

[Smoller, 1983] J. Smoller, \textit{Shock Waves and Reaction-Diffusion
Equations,} Springer-Verlag, New York, 1983.

[Soskin, 1998] M. S. Soskin (ed.), \textit{Singular Optics,} \textit{Proc.
SPIE,} vol. 3487, Optical Society of America, Washington, 1998.

[Stoker, 1947] J. J. Stoker, The formation of breakers and bores, \textit{%
Commun. Pure \& Appl. Math.} \textbf{1} (1948), 1-87.

[Stoker, 1957] J. J. Stoker, \textit{Water Waves}, Interscience, New York,
1987.

[Stramma \& Lutjeharms, 1997] L. Stramma and J. R. E. Lutjeharms, The flow
field of the subtropical gyre in the South Indian Ocean into the Southeast
Atlantic Ocean: a case study, \textit{J. Geophysical Res.} \textbf{99}
(1997), 14053-14070.

[Tricker, 1965] R. A. R. Tricker, \textit{Bores, Breakers, Waves, and Wakes,}
American Elsevier, New York, 1965.

[White \& Fornberg, 1998] B. S. White and B. Fornberg, On the chance of
freak waves at sea, \textit{J. Fluid Mech.} \textbf{355} (1998), 113-138.

[Whitney, 1955] H. Whitney, On singularities of mappings of Euclidean
spaces, I. Mappings of the plane into the plane, \textit{Annals of Math.} 
\textbf{62} (1955), 374-410.

[Wu, 1997] S. Wu, Well-posedness in Sobolev spaces of the full water wave
problem in 2D, \textit{Invent. Math.} \textbf{130} (1997), 39-72.

[Yoshihara, 1982] H. Yoshihara, Gravity waves on the free surface of an
incompressible perfect fluid of finite depth, \textit{Publ. Res. Inst. Math.
Sci.} \textbf{18} (1982), 49-96.

\end{document}